# SURVEY ON MODELLING METHODS APPLICABLE TO GENE REGULATORY NETWORK

Chanda Panse[1] and Dr. Manali Kshirsagar[1]

[1]Department of Computer Technology, Yeshwantrao Chavan College of Engineering, Nagpur, India

[1]Department of Computer Technology, Yeshwantrao Chavan College of Engineering, Nagpur, India

## ABSTRACT

*Gene Regulatory Network (GRN) plays an important role in knowing insight of cellular life cycle. It gives information about at which different environmental conditions genes of particular interest get over expressed or under expressed. Modelling of GRN is nothing but finding interactive relationships between genes. Interaction can be positive or negative. For inference of GRN, time series data provided by Microarray technology is used. Key factors to be considered while constructing GRN are scalability, robustness, reliability and maximum detection of true positive interactions between genes. This paper gives detailed technical review of existing methods applied for building of GRN along with scope for future work.*

## 1. INTRODUCTION

Cellular biology is now becoming growing area for researchers to carry out research. There is ample amount of biological data available because of different, advanced experimental technologies like Microarray. In order to analyze and retrieve informative knowledge from these data efficient computational methods are required. These methods helps in knowing interactions carried out in organisms at genomic levels through some mathematical formalism. Such gene to gene communications are represented in terms of special network known as Gene Regulatory Network (GRN). It is a graphical representation in which nodes consist of genes or protein or metabolites and edges connecting them show regulatory relationships between them. It is constructed by observing the behaviour of genes and their impact on other genes at particular experimental condition. This behaviour is analysed in terms of measuring the values of gene expressions with the help Microarray technology. Gene expressions are dynamic and non-linear in nature.

When specialised proteins known as Transcription Factors (TFs) bind to promoter region of DNA, it intervenes the rate of protein synthesis which in turn results in sudden change in values of gene expressions. This intervention can be positive or negative. If rate of protein synthesis increases then it is called as activation or up-regulation or positive regulation and if it decreases then it is called as inhibition or down regulation or negative regulation. If gene g1 positively regulates gene g2 then it mean there is increase in expression level of g2 because of g1 and in negative regulation it decreases expression level of g2. GRN is useful for analyzing the effect of drugs on genes. It also helps in studying dynamics of specific gene under particular diseased or experimental conditions. It helps in studying diseases caused due to genes (hereditary diseases). Considering all advantages of GRN, identifying GRN is quite complicated and interesting task for researchers. There are many mathematical models like [1][2][6][26][17][13] proposed for inference of GRN but still this research area is in the stage of babyhood because of inability to reach to maximum detection of true positive interactions between genes.

Depends on the availability of models they are classified into two major categories i.e. traditional and untraditional. A traditional model includes Boolean Network, Bayesian Network, Linear Differential Model etc and non-traditional model includes Neural Network model and Model based on Evolutionary algorithms. This paper gives detailed review of all those mathematical formalisms used for identifying GRN along with database and experimental setup used. It also describes challenges while handling gene expression data generated from microarray technology.

The paper is organized as follows: section 2 describes properties of GRN. Section 3 talks about difficulties in gathering data. Section 4 discusses detailed literature survey of existing models in terms of model name, mathematical representation and database used accompanying with advantages and limitation followed by summary and conclusion in last section.

## 2. PROPERTIES

Gene regulation occurs at any stage of transcription, translation and post-transcription or post-translation. Regulatory elements involved can be proteins (TFs) or genes or RNAs or metabolites. Depends on elements involved and mathematical model used GRN has following properties which have to be considered while inventing new method [11].

### 2.1. Anatomy of network

GRN is represented using graph. Anatomy of GRN is nothing but the topology of how regulatory elements are arranged in it. Generally genes are used as nodes of graph and solution from mathematical models decoded into directed or undirected edges depends on models used. Those edges can be activatory or inhibitory. How to represent genes and different types of interactive edges is going to create topology of GRN.

### 2.2 Sensitivity

GRN are insensitive to the variations in model parameters. This robustness is guaranteed by models following specific topology. Robust nature of GRN assures same plan for functioning of genes even though genome is dynamic in nature.

### 2.3 Noise

Data required for modelling GRN is a biological data which has inherited noise from biochemical reactions. This noise can be advantageous or disadvantageous to some models. Therefore handling this noise is the main concern. It is considered in terms of inclusion of some parameters in mathematical form of model

## 3. CHALLENGES

For the designing of GRN time series data from Microarray is used which is in the form of matrix of order N by M where N represents number of genes and M represents length of experiments which is shown in Fig1. Gij represents expression of gene i at jth time interval of particular experimental condition.

| Genes | $T_1$ | $T_2$ | ....... | $T_M$ |
|-------|-------|-------|---------|-------|
| $G_1$ | $G_{11}$ | $G_{12}$ | ....... | $G_{1M}$ |
| $G_2$ | $G_{21}$ | $G_{22}$ | ....... | $G_{2M}$ |
| : | : | : | : | : |
| : | : | : | : | : |
| $G_N$ | $G_{N1}$ | $G_{N2}$ | ....... | $G_{NM}$ |

Figure 1. Microarray time series data

As microarray captures temporal response of genes it is very difficult to select specific model which will analyze these expression patterns of genes and state causal relationship among them. Because of this there are many computational challenges described by [29] at different levels of analysis of time series gene expression. Those four different levels are experimental design, data analysis, pattern recognition and network. This section is going to focus on challenges to be faced at above levels.

If there is deficiency in sampling (under sampling) then there is a possibility of missing major change in gene expression which might respond in later stage of sampling. Along with this if ample amount of samples are needed then again it is very time consuming as well as costly process. This is the main challenge at the level of experimental designing.

While doing data analysis there is possibility of unavailability of continuous gene expression for every time slot. This is the challenge at second level. Microarray experiments also add noise to data which might or might not be advantageous. For calculating missing gene expression some effective interpolation techniques are needed.

Third challenge is at pattern analysis level. In pattern analysis genes are grouped into different clusters on the basis of their mere expression values. Built in dependencies of consecutive time points are not exploited by existing clustering algorithms which is very important in GRN modelling. None of the clustering algorithms is taking advantage of this while clustering.

Major challenge at network level is going to generation of such a predictive model which actually predict biologically possible interaction among genes. i.e. which gene is going to interact with which other gene, which gene is going to expressed first, which gene is going to expressed next. Thus considering limitations of sampling and data analysis it requires pre-processing of gene expression data before designing model. Pre-processing is in terms of clustering or noise removing or interpolating missing values. Next section is going to describe methods and algorithms proposed by different authors for modelling GRN.

## 4. MATHEMATICAL MODELS

There are large number mathematical models proposed for inference of GRN. On the basis of their chronological order those are classified into: traditional and un-traditional models. Traditional model includes Boolean network, Bayesian network, Dynamic Bayesian. network model, Differential equation model and untraditional model includes models based on evolutionary algorithms like Genetic algorithms and PSO.

The model discussed in [9] is based on Boolean network without time delay. Value of gene regulatory relationship is calculated by applying Boolean function which contains Boolean operators like AND, OR, NOT etc. on at most two genes contributing to in degree of some other gene G. This model considers conditions of gene disruption and over expression for building GRN but it is limited to only two regulators not more than that.

Discrete time model developed in [20] is based on weight matrices. All regulatory interactions between genes are computed as linear coefficient which is summation of independent regulatory input. i.e. net regulatory effect of gene j on gene i at some time t is given by the expression

$$r_i(t) = \sum_j w_{i,j} \, u_j(t)$$

where $W_{ij}$ is regulatory influence of gene j on i and $u_j(t)$ represents expression level of gene j at time t. Positive value of $W_{ij}$ represents activation of gene i by gene j and negative value indicates inhibition. This model suffers from data loss because of not considering dependency in gene interactions.

Hybrid model developed in [30] uses integration of Boolean network and S-system model. Boolean network of this model helps in treating large genome data and S-system model adds loop structure in genetic network which is important and not discussed in prior existing models. From the observation of pattern of gene expressions a new model is proposed in [12]. From those observation directed graph with labelled edges are constructed. Because of possibility of existing false positive regulatory edges the graph is then pruned using combinatorial optimization problem i.e. simulated annealing but concluded with assumption that some factors like phosphorylation are required for activation of gene expression which cannot be measured by gene chips at RNA levels.

Because of only considering two states of gene by Boolean network, intermediate state get neglected [25] [31]. As an extension of Boolean network, Generalized Logical Network (GLN) is developed by Thomas and colleagues [32]. It allows having intermediate values of gene expression (other than 0 and 1) and transition can occur synchronously and asynchronously between these values. In GLN each node (gene) is associated with generalized truth table (GTT) which contains all possible values of parent nodes. This GTT maps values of parent nodes to the value of corresponding child node.

Because of complex nature of gene regulatory process, model based on Bayesian Network (BN) is proposed in [33]. BN has ability to capture complex stochastic process. In this model genes are assumed to be random variables and regulatory interaction is obtained by applying joint probability distribution for each random expression of gene. The Bayesian network which formed, gives the topology of GRN. Drawback of this network is that it fails to consider dynamic nature of genes as well feedback loops. To overcome this the model is extended to Dynamic Bayesian Network(DBN) which considers temporal aspects to gene data.

Inference of GRN using DBN is same as that of Bayesian network except that extra parameters of time t-1 are added [35]. It is limited to small network because of high computational overhead. Another two probabilistic graphical models used for reconstruction of GRN using microarray are discussed in [19]. One is independence graph model which is based on forward and backward search algorithm and another is based on Gaussian Network (GN) model which is rooted on Greedy Search method. Out of these two models GN provides better results in terms of gene-gene interactions than later model but it suffers from longer learning time.

These are all traditional models which can either applied to discrete or continuous data.

In [24] author proposed Neural network based model. Genes are represented by nodes of neural network and connection between nodes represents regulatory relationship between genes and

16

weight matrix obtained after training gives type of regulatory relationships. Expression of a gene $g_i$ at time t+Δt is obtained from weight of edge $W_{ij}$ between nodes i,j and expression $y_i$ at time t. Thus regulatory effect on gene $g_i$ is obtained from

$$g_i \approx \sum_j w_{ij} y_j$$

Sigmoid function is then applied to this equation to transform it in the interval <0,1>. The model was computationally intensive because of large number of parameters obtained from experimental data. This is the first non-traditional model used for inference of GRN. It provides better description of biological system and can be used to better measure the reasonableness of mathematical model.

Another model based on linear differential equation is implemented in [21]. In this model author has utilized a statistical method the Lasso to reverse engineer the network. The linear time-continuous dynamical model is of the form

$$\frac{dx_i(t)}{dt} = \sum_{j=1}^{N} w_{ij} x_j(t) + \epsilon_i$$

This method helps in providing insight of GRN on a large scale but cannot be used to get detailed information of small sub-networks related to biological function.

In order to deal with uncertainty in the function of genes, models based on fuzzy system are described in [16][5][8]. Once clustering is done in [5], the model is simplified as linear model and with the help of evidence code generated by Gene Ontology. This assessed function of gene and gene interaction. But the major limitation of this method is that it shows relationship between clusters instead of individual gene. Another fuzzy based model in [16] talks about fuzzy data mining technique which transfers quantitative gene expression values into linguistic terms which in turn discovers fuzzy dependency relationships between genes from unseen samples and genes in original database.

 Dynamic fuzzy model is developed in [8]. It incorporates structural knowledge in gene network. This helps in knowing which gene affects another gene and whether effect is positive or negative. It achieves fast identification of regulatory interaction and captures inherent non-linear dynamic property of gene network so that prediction of fuzzy gene network will be easier.
 A two stage methodology based on Genetic Algorithm (GA) and Expectation Maximization algorithm is implemented in software GnetXp is mentioned in [10].  In first stage gene expression trajectories are clustered based on expression values and in second stage Kalman filtering is applied to parameters of first order differential equation which describes dynamic of GRN in terms of gene interaction and future time points prediction.

 Another method based on ordinary differential equation is employed in [15] which developed mathematical formulation for identifying gene interaction from time series expression profiles. But the method is only applicable to small network consisting of 4-8 genes Similar to [10] a method is proposed in [4]. In this model instead of GA, genetic programming is used to infer ordinary differential equation framework with noise using time-series data and Kalman filtering is used to estimate model parameters. Considering this method as a base, Extended Kalman Filtering (EKF) approach is used to model non-linear dynamics of GRN in [3]. This method helps to deal with scarcity of time series gene data and EKF is applied to calculate parameters of model estimated by differential equation.

As Bayesian Network based modelling is time consuming hybrid approach is proposed in [7]. It is based on gene expression and gene ontology. This method is going to detect potential gene regulatory pair which cannot be detected by any general inferring method and also identifies feedback loops between genes. For this purpose overlap clustering is used to find local components of global network so that search space of Bayesian Network method get reduced and then BN is applied on each cluster to infer local networks.

The algorithm based on Bayesian network is proposed in [17]. This algorithm performs several independence tests on given biological data and then build network from it. Although the algorithm is for sparse graph, it identifies dense nodes first and then constructs network. The main advantage of this method is that it can also applicable for large network and helps in achieving good quality network with minimum computational overhead.

Method discussed in [18] overcomes the problem of Bayesian Network of neglecting time series data which is useful in identifying direction of regulatory interactions among genes. This method uses the concept of influence vector. Value of this influence vector associated with individual gene increases as gene expression increases with time step. Finally vector with highest score helps in building GRN.

In order to overcome scalability issue of previous methods Markov blanket method for modelling GRN is presented in [2]. The method starts with formation of Markov Blankets for each gene X. Putative network model is selected using first and second order partial co-relations on the basis of achieved excellent meaningful relationships. Again false positive edges are removed from remaining model using guided GA on the basis of overall network score. Disadvantage of this method is that there is loss of data as it is applicable to discrete data. Advantage is that it is fast because of calculation of partial correlations only up to second order and also because of guided GA which keeps track of intermediate results and calculations.

Another class of evolutionary algorithms i.e. Particle Swarm Optimization (PSO) is used for finding interaction among genes in many models [26][14][22][6].

Considering limitations of all traditional modelling methods, the best formalism to tackle complex time series gene data is Recurrent Neural Network (RNN). Inference of GRNs with RNN Models using PSO uses PSO to train parameters of RNN based model and type of regulatory interaction between genes is stated with the help of weight matrix obtained after training.

Modelling GRN using hybrid DEPSO [26] states that main difficulty with RNN is its training. Training of RNN is done with three different methods: Differential Evolution (DE), PSO, and hybrid of these two (DE and PSO) DEPSO. Out of these three DEPSO proved to be efficient than individual method which is then applied for inference of GRN. Parameters of the model are calculated by training of RNN using this hybrid approach. It has been observed that the method provides meaningful insight in capturing non-linear dynamics of gene networks and reveals gene to gene interaction through weight matrix.

Reverse Engineering of Gene Regulatory Networks using Dissipative Particle Swarm Optimization [22] applied PSO for calculating parameters of S-system which creates mathematical framework for reverse engineering of GRN. S-system is a Biochemical system theory represented by following mathematical equation:

$$\frac{dX_i}{dt} = \alpha_i \prod_{j=1}^{N} X_j^{g_{ij}} - \beta_i \prod_{j=1}^{N} X_j^{h_{ij}}$$

Kinetic order $g_{ij}$ regulates activation of gene expression $X_i$ due to $X_j$ and $h_{ij}$ regulates inhibition of $X_i$ due to $X_j$. The method helps in minimizing false positive interactions. It is specifically applicable to small network of 5-8 genes.

Swarm Intelligence Framework for Reconstructing Gene Networks [6] uses computationally intelligent two swarm intelligence algorithms i.e. PSO and Ant Colony Optimization (ACO) for retrieving biologically possible architecture along with RNN. Different architectures are generated using RNN whose parameters are again trained by PSO and ACO is used for reducing searching time of networks using stochastic graph generation.

Thus many methods have been reviewed which are summarized in table 1. In accompanying with this, advantages, limitations, and modelling equations of Bayesian Network, Boolean network, ordinary differential equation model, and neural network model are discussed in detailed in [35].

Table 1. Summary Table.

| Ref. paper | Experimental Database | No of Genes | Methodology Used | Findings |
|---|---|---|---|---|
| [1] | Heat stress response in Drosophila | 36 time points with 360 genes | State Space Model with Kalman Filter | Genes have non-linear dynamic nature and GRN is generally sparse |
| [2] | Spellman's Microarray | 800 | Markov blanket with GA is used | Method is robust with low false discovery rate also satisfies scalability issue |
| [4] | Yeast protein Synnthesis + synthetic data | 17 time points with 12 | Genetic Programming with Kalman filter | Factors like noise and perturbation have effects on gene expression |
| [5] | Carbohydrate Metabolism in Arabidopsis Thaliana | 41 time points with 484 genes | Fuzzy Clustering with GO annotations | Presents the regulatory relationship between gene cluster which is of less interest to researchers, feedback loops are present in GRN which is positive finding |
| [6] | E coli | 50 time points after every 6 minutes with 8 genes | Swarm Intelligence with RNN | Achieves biologically plausible interaction to better extent |
| [10] | Human fibroblast genome | 12 time points with 517 genes | GA with Kalman filtering | Kalman filter helps in handling noise and missing and irregularly space data |

| [12] | Cell division in S. Cerevisiae | 17 time points with 6601 genes | Clustering along with simulated Annealing | A gene can be regulated by multiple transcriptions not limited to two only |
|---|---|---|---|---|
| [13] | Artificial data | 10 time points with 10 genes | GA | For accurate inference static as well dynamic analysis is needed |
| [15] | Yeast cell cycle | 17 genes | Broyden–Fletcher–Goldfarb–Shanno (BFGS) method of optimization | Method considers time delay and noise and is applicable to small network |
| [16] | S. Cerevisiae | 800 genes | Fuzzy data Mining technique | Reveals biologically meaningful relationship |
| [17] | Yeast and human Raf Pathway | 17, 25 genes | PC algorithm | Mixing of more than one type of data improves quality of network |
| [18] | S. Cerevisiae | - | GeneNet algorithms | Overcome problems of Bayesian Network but Suffers from problem of local minima |
| [19] | MAPK pathway in yeast | 46 time points with 6221 genes | Independence graph with forward or backward search algorithm and Gaussian network with greedy search method | GN provides better result in terms of gene-gene interaction but suffers from long learning time |
| [20] | Artificial database | - | TReMM (Transciption Regulation Modeled with Matrices) algorithm | Assumes that genetic interactions are independent, model suffers form loss of data and some contradictory regulatory pathways |
| [22] | E. Coli | | Particle Swarm Optimization | PSO works better in term of recognition of correct gene-gene interaction in noisy data, with increase in population model converge fast with increase computation |
| [23] | Artificial data | 3 genes | Neural Network | Provides better description of biological system because of stochastic nature of neural network |
| [26] | E. coli database with different light intensities | 50 time points after every 6 minutes with 8 genes | differential evolution with PSO | Hybrid approach helps in identifying Genes with similar expression which helps in reducing false positive interaction between genes |

| [34] | Rat gene database under different experimental conditions | 28 time points with 65 genes | Linear Additive model | Model is robust and simple to implement and can be used for large data set |
|------|------|------|------|------|

## 3. CONCLUSIONS

Thus, this paper has discussed many methodologies used for inference of GRN. Many are limited to small GRN consisting of 5-8 genes. Many methodologies are able to find more true positive interactive edges but along with that they also added false positive edges in the network. None of the method also discussed about how to deal with curse of dimensionality problem. Many traditional models are focussing on either structure or structure along with dynamics [28]. There is lot more scope in modelling GRN with respect to scalability, robustness and reliability. Major source of Database required for GRN is Gene Expression Omnibus (GEO) which is freely downloadable. In addition to this tools along with type of data handled by particular method are discussed in [27].

**Authors**

**Chanda Panse** received her Bachelor of Engineering degree from Pune University in 2004. She has completed her M.E. in Computer Science and Engineering from BITS Pilani, Rajasthan in 2008. She is an Asst. Professor in Yeshwantrao Chavan College of Engineering, Nagpur. She has around 5 Yrs of teaching experience. Currently she is persueing her Ph.d in Gene Regulation from Nagpur University. Her area of research includes Data Structures, Operating Systems,  Parallel Programming and Bioinformatics. 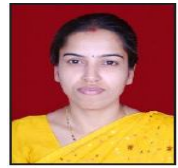

**Dr. Manali Kshirsagar** received her B.E. in Computer Technology, M.E. Computer Science & Engineering and Ph.D. in the faculty of Computer Science and Information Technology. Her specialization at the PG and Ph.D. level has been in the areas of Data Warehousing , Data Mining and Bioinformatics. She has about 20 technical and research publications to her credit which have appeared in International Journals, International and National conferences. She is guiding 08 Ph.D. students at present. With a total experience of 18 years of teaching and industry. She was  Head of Computer Technology Department of YCCE ,Nagpur from 2008 till 2012. She is currently a Vice- President of ADCC Infotech, Nagpur. 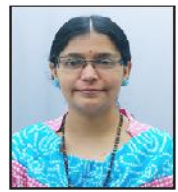